# Gravity-Awareness:
# A Deep Learning and LLM Framework for Predicting Human Adaptation to Altered Gravitational Environments

Bakytzhan Alibekov [1], Alina Gutoreva [1*] & Elisa Raffaela Ferre [2]

1. School of IT & Engineering, Kazakh-British Technical University, 050000, Tole bi 59, Almaty, Kazakhstan.

2. School of Psychological Sciences, Birkbeck University of London, Malet St, London WC1E 7HX, United Kingdom.

* Correspondence to:
Alina Gutoreva, PhD, a.gutoreva@kbtu.kz, Kazakh-British Technical University, Almaty, Kazakhstan, 050000

**ORCID**
Bakytzhan Alibekov: https://orcid.org/0009-0006-7245-1179
Alina Gutoreva: https://orcid.org/0000-0001-6893-7252
Elisa Raffaela Ferre: https://orcid.org/0000-0002-0643-848X

**Funding Note.** This research received no specific grant from funding agencies in the public, commercial, or not-for-profit sectors.

**Conflict of Interest Statement.** The authors have no other instances to declare that could be perceived as a possible conflict of interest.

**Data Availability Statement.** Data is available on Zenodo by the link: https://doi.org/10.5281/zenodo.17450642

**Abstract**

Earth's gravity fundamentally shapes human behaviour. The brain encodes this force as an *internal model of gravity*, enabling the prediction and interpretation of gravitational effects during perception and action. Understanding how this model adapts to altered gravity is critical for predicting human performance in spaceflight. We present a computational framework for modelling neurophysiological adaptation across diverse gravitational environments. The framework has two components trained on open-access data from altered-gravity studies, particularly parabolic flights. The first component (CorticalG) employs a lightweight multilayer perceptron neural network to predict gravity-dependent changes in EEG frequency bands, estimating cortical state under different gravitational loads. The second component (PhysioG) uses independent Gaussian process models to capture broader physiological responses, including heart rate variability, electrodermal activity, and motor control. To complement the quantitative modelling, we simulated subjective experience across gravitational environments using the Large Language Model (LLM) Claude 3.5 Sonnet. Physiological outputs prompted the model to generate narratives describing alertness, bodily awareness, and cognitive state across zero gravity, partial gravity of the Moon and Mars, and hypergravity. This framework provides a novel approach for investigating human adaptation to spaceflight. It offers a predictive tool to assess performance and resilience, supporting the design of future space exploration missions.

## 1. Introduction

Human behaviour has been fundamentally shaped by Earth's constant gravitational force, 9.81 $m/s^2$, commonly referred to as 1g. Gravity provides a pervasive sensory signal that the brain integrates to predict and interpret interactions with the environment. This integration involves the integrated processing of vestibular, visual, and proprioceptive inputs, allowing the brain to maintain situational comprehension, improve perception, guide action and support cognitive function. Understanding how this internal representation adapts when gravity deviates from Earth's standard 1g, whether in microgravity, the partial gravity of the Moon (0.16g) or Mars (0.38g), or during hypergravity, is critical for anticipating human performance, resilience, and safety during upcoming crewed lunar and deep-space missions. To address this, we developed a computational framework for modelling neurophysiological adaptation across diverse gravitational environments. The framework consists of two complementary components. The first employs a lightweight multilayer perceptron neural network to predict gravity-dependent changes in EEG activity, providing an estimate of cortical state under different gravitational loads. The second uses independent Gaussian process models to capture broader physiological responses. To extend these quantitative predictions to subjective experience, we simulated introspective narratives across gravitational conditions using the Large Language Model (LLM) Claude 3.5 Sonnet. Physiological outputs from the computational framework prompted the model to describe perceived alertness, bodily awareness, and cognitive state under zero gravity, the partial gravity of the Moon and Mars, and hypergravity during spacecraft launch and re-entry.

### 1.1 From Earth to Space: Neural Integration and Adaptation to Altered Gravity

The human nervous system does not rely on a single receptor to encode gravity; rather, it integrates information from multiple sensory modalities to construct a coherent representation of gravitational forces (Angelaki & Cullen, 2008). On Earth, the vestibular otolith organs of the inner ear detect linear acceleration and head tilt relative to gravity, providing a foundational signal about its direction (Goldberg et al., 2012). These vestibular inputs are integrated with visual cues, such as optic flow and horizon orientation, and proprioceptive feedback from muscles and joints, enabling the brain to maintain a stable perception of the body and its environment (Merfeld et al., 1999). This multisensory integration gives rise to the *internal model of gravity*, a dynamic neural representation that allows organisms to predict and interpret gravitational forces while planning and executing adaptive behaviours (Lacquaniti et al., 2014).

From a predictive-processing perspective, the body and environment form a tightly coupled system: the body continuously shapes the environment it acts within, while the environment constrains and informs bodily and cognitive regulation (Clark & Chalmers, 1998). Gravity provides a constant reference for sensorimotor and perceptual predictions. Everyday actions, such as catching a ball, depend on the brain's ability to integrate vestibular, visual, and proprioceptive signals under the assumption of 1g to accurately anticipate trajectories and guide movements. When gravity is altered, as during spaceflight, this coupling is disrupted, forcing simultaneous recalibration of bodily and environmental predictions (Hohwy, 2013; Apps & Tsakiris, 2014). Studying these conditions informs the development of countermeasures for deep-space missions and provides a window into embodied cognition, revealing how the brain constructs a unified model of self within the environment (Gawron, 1997).

Deviation from Earth's standard 1g during spaceflight provokes both structural and functional changes in the human brain. Post-flight MRI studies have documented macrostructural shifts, including an upward displacement of the brain within the skull and narrowing of the central sulcus (Roberts et al., 2017; Lee et al., 2019), reflecting adaptations to chronic microgravity-induced fluid redistribution. Microstructural alterations are also evident, with reductions in grey matter volume in frontal and temporal regions alongside increases in sensorimotor areas (Koppelmans et al., 2016; Hupfeld et al., 2020a), likely reflecting functional reweighting of neural circuits to accommodate the altered sensorimotor demands of microgravity. White matter changes have been observed in tracts supporting vestibular and proprioceptive processing (Doroshin et al., 2022), suggesting that communication between cortical and subcortical regions is remodelled in response to persistent changes in gravitational input. Additionally, ventricular enlargement has been consistently reported post-flight, linked to cerebrospinal fluid redistribution and transient increases in intracranial pressure (Van Ombergen et al., 2019a; Kramer et al., 2020).

Functional neuroimaging indicates that exposure to gravitational deviations induces widespread reorganization of brain network dynamics. Vestibular and sensorimotor networks show decreased coupling, reflecting reduced reliance on gravity-dependent inputs, while visual and proprioceptive systems exhibit increased engagement to compensate for altered spatial cues (Pechenkova et al., 2019; Hupfeld et al., 2022). Altered connectivity is also observed between the cerebellum, motor cortex, and default mode network (Demertzi et al., 2016; Van Ombergen et al., 2017a), suggesting integrated modulation of sensorimotor and higher-order cognitive networks under non-1 g conditions. Concurrently, deactivation in somatosensory and visual cortices further supports a functional reweighting of sensory contributions, illustrating the brain's adaptive strategies to maintain successful behaviour in altered gravity environments.

Electrophysiological evidence further highlights functional adaptations: alpha-band (8–13 Hz) power and connectivity within the Default Mode Network (DMN) are reduced during and after spaceflight, indicating diminished cortical inhibition and reflective processing (Pusil et al., 2023). Conversely, beta-band (12–30 Hz) activity increases in frontotemporal and sensorimotor regions during microgravity, reflecting compensatory activation to maintain posture and body schema (Quivira-Lopesino et al., 2025). These shifts reflect a reorganisation of cortical dynamics under microgravity: reduced DMN activity suggests diminished resting-state self-referential processing, while increased frontotemporal and sensorimotor beta activity indicates compensatory engagement of cognitive control and postural stabilisation networks (Klimesch, 1999).

## Gravity-Awareness: A Novel Integrative Framework for Neural Adaptation during spaceflight

We propose the term "gravity-awareness" to describe the human organism's integrated, system-level response to gravity[1], encompassing neurophysiological, sensorimotor-behavioural, and perceptual-cognitive domains. The neurophysiological component reflects the brain's processing of gravitational input, including cortical oscillations, network connectivity, and autonomic signals such as EEG band-power changes, heart rate variability, and electrodermal activity. These neural processes support the sensorimotor-behavioural domain, which includes postural adjustments, locomotor control, and coordinated motor actions that allow the body to interact effectively with its environment under varying gravitational loads. Perceptual-cognitive processes, in turn, rely on the integrity of both neural processing and sensorimotor function, encompassing higher-order abilities such as spatial orientation, situational awareness, working memory, decision-making and executive function, all of which depend on a stable gravitational reference to guide perception and action accurately.

[1] Gravity, at Earth's surface, is fundamentally an acceleration of 9.81 $m/s^2$ rather than a force in itself; the gravitational "force" experienced by organisms is the contact reaction opposing free-fall. Consequently, the nervous system responds not to gravity per se but to the inertial consequences of acceleration (Hartle, 2003). This distinction is critical for understanding human adaptation to altered gravitational environments: physiological and cognitive responses to artificial linear accelerations, as experienced in centrifuge or parabolic flight experiments, closely parallel those elicited by true changes in gravitational field strength.

These domains of gravity-awareness are therefore tightly interdependent, forming a dynamic system in which changes at one level influence others. For example, neurophysiological alterations such as suppression of alpha-band activity in microgravity can constrain cognitive processing, which in turn affects sensorimotor performance. Conversely, disruptions in posture, locomotion, or fine motor control can feed back to modulate attentional and executive resources. Gravity-awareness therefore reflects the degree to which an individual's integrated neurophysiological, sensorimotor, and cognitive functioning remains adaptive across different gravitational environments (Arshad & Ferré, 2023; Lacquaniti et al., 2014). This framework consolidates previous concepts, including gravity perception, gravitational orientation, and embodied spatial awareness, into a unified operational model suitable for computational modelling and predictive assessment of human performance during spaceflight.

Empirical evidence demonstrates that exposure to altered gravitational environments profoundly affects sensorimotor function, often manifesting as impaired postural control, disrupted locomotion, degraded fine motor coordination, and balance instability (Arshad & Ferré, 2023; Schneider et al., 2008; Dev et al., 2024; Manzey & Lorenz, 1998). These sensorimotor deficits frequently give rise to secondary perceptual disturbances, including spatial disorientation, altered spatial scaling, and motion sickness, reflecting the central role of vestibular input in constructing a coherent body-in-space representation. The modulation of gravity-awareness under these conditions arises from the interplay of sensory conflict, neurovascular alterations, and neuroplastic adaptation. Sensory conflict emerges when vestibular signals no longer align with visual or proprioceptive information, particularly in micro- or hypergravity, degrading the integration of multisensory cues and impairing the brain's ability to maintain a stable spatial framework (Bles et al., 1984; Lackner & DiZio, 2006). Neurovascular changes, including shifts in fluid distribution, intracranial pressure, and cerebrovascular perfusion, further influence neuronal signalling and cortical processing. Computational and empirical studies indicate that microgravity increases hemodynamic pulsatility in cerebral microcirculation, suggesting a direct mechanism by which reduced gravitational loading can modulate neural activity and functional connectivity (Scarsoglio et al., 2023; Roberts et al., 2017). Concurrently, neuroplasticity enables structural and functional adaptation across sensorimotor and cognitive networks, facilitating partial compensation for altered input; however, rapid or extreme gravitational transitions may exceed the nervous system's immediate compensatory capacity, particularly during early exposure phases (Van Ombergen et al., 2017; Reschke et al., 1998).

These sensorimotor and neurophysiological perturbations propagate upward to affect higher-order cognitive functions. Microgravity has been shown to impair voluntary attentional control in the visuospatial domain and to slow processing speed during complex cognitive tasks (Salatino et al., 2021). Over longer-duration missions, cognitive capacities such as sustained attention, working memory, and spatial orientation are particularly vulnerable to the cumulative effects of gravitational and non-gravitational stressors, including isolation, confinement, and circadian disruption, highlighting the interdependence of sensorimotor integrity and higher cognitive performance (Faerman et al., 2023; Clément & Ngo-Anh, 2013).

Gravity-awareness provides a unifying conceptual framework that links these neurophysiological, sensorimotor, and cognitive processes, capturing the mechanisms through which the human organism detects, interprets, and adapts to changes in gravitational load. By emphasizing the continuous feedback between body, brain, and environment, this framework accounts for both the acute disruptions and adaptive plasticity observed under altered gravity. It offers a rigorous basis for experimental investigation, operational monitoring, and the design of countermeasures in spaceflight, highlighting the remarkable resilience and flexibility of human neural and behavioural systems when confronted with profound changes in gravitational input.

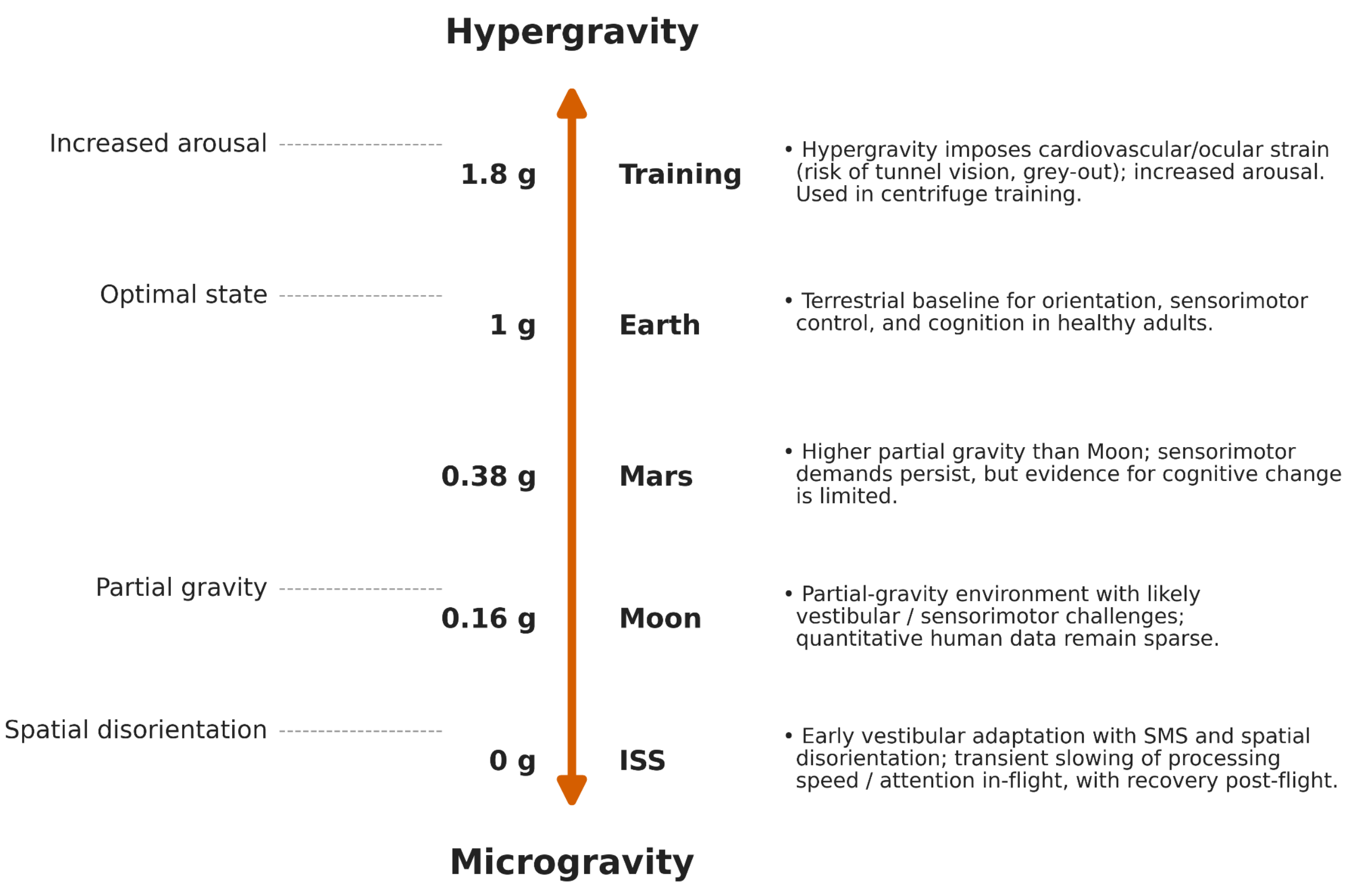

**Figure 1. Gravity-awareness and is cognitive–behavioural effects.** The schematic illustrates gravity levels and their associated cognitive and physiological effects. Under microgravity (0g), individuals commonly experience spatial disorientation, impaired spatial memory, and space motion sickness during early adaptation phases. In partial-gravity environments (≈0.16–0.38g), vestibular cues are suboptimal, resulting in slowed motor coordination and increased reliance on visual feedback. At 1g, terrestrial conditions provide the baseline for optimal orientation, sensorimotor integration, and cognitive performance. Under moderate hypergravity (≈1.8g), increased cardiovascular load and fatigue are accompanied by temporary reductions in working memory and executive performance. *Note:* Current evidence on long-term human performance in partial-gravity is indirect; experimental and modelling studies suggest that gravity levels below approximately 0.4g may be insufficient to maintain musculoskeletal and cardiopulmonary function without countermeasures (Clément et al., 2024).

## Computational and Machine Learning Models of Gravity-Awareness

Gravity-awareness can be formalized computationally as the brain's inference over a latent gravity vector, integrating multisensory input to maintain a coherent representation of "upright". Within a Bayesian framework, noisy vestibular, proprioceptive, and visual signals act as likelihoods, while prior expectations encode Earth's 1 g and an egocentric "down." Because this prior reflects years of accumulated experience, it typically dominates individual sensory cues, maintaining perceptual stability even when sensory signals are transiently inconsistent. This approach accounts for several characteristic behaviours observed in altered-gravity environments, including slow adaptation during prolonged exposure to microgravity or partial gravity, hysteresis when transitioning between gravitational states, and dynamic reweighting of sensory cues according to their relative reliability, for example, reductions in vestibular noise increase visual dominance in spatial orientation (Körding & Wolpert, 2004; Zago & Lacquaniti, 2005).

Computational approaches, such as Kalman filtering, have long captured vestibular-sensory integration. By fusing canal-otolith dynamics, these models disambiguate tilt from linear translation and incorporate velocity storage mechanisms (Merfeld et al., 1999; Laurens & Droulez, 2007). Extending the Kalman filter to include a gravity state allows construction of a gravity-aware estimator, in which internal predictions derived from body dynamics are continuously corrected by sensory residuals. Under altered gravitational conditions, the assumed process model is biased, and adaptation corresponds to slow recalibration of internal gravity estimates and noise parameters, consistent with the gradual behavioural and perceptual adjustments observed in altered gravity environments.

Building on these computational principles, we adopt a dual modelling approach that combines classical machine learning (ML) with LLMs to simulate human gravity-awareness. Recent advances in LLMs demonstrate their potential to function as in silico participants capable of generating structured, biologically plausible responses to complex cognitive and physiological scenarios. We leveraged an advanced LLM (Claude 3.5 Sonnet) to simulate subjective awareness under varying gravitational states. The model was prompted with a structured scaffold of physical, physiological, and sensorimotor parameters, including gravitational load, head orientation, postural sway, gait dynamics, muscle activation, and autonomic signals such as heart rate variability. These constraints allowed the LLM to produce plausible cognitive and embodied responses, capturing alertness, self-awareness, and sensorimotor experiences that are otherwise difficult to measure directly under extreme or ethically constrained conditions. While prior studies have shown that LLMs can emulate aspects of human cognition such as moral reasoning or social trait judgments (Levy et al., 2023; Ullman, 2023), their application to simulating embodied awareness in altered gravity is novel.

The present study is structured in two parts. Part 1 introduces two complementary ML models that quantify human gravity-awareness. The first model focuses on neural activity, predicting continuous, g-load-dependent changes in key electroencephalographic (EEG) frequency bands, which reflect cortical processing. Implemented as a Multi-Layer Perceptron (MLP) enhanced with Fourier features, it captures the nonlinear relationship between gravitational load and cortical network activity, including networks such as the Default Mode Network (DMN) and the Prefrontal Cortex (PFC). The second model addresses broader physiological responses, predicting autonomic nervous system activity and motor behaviour. Built using a suite of independent Gaussian Processes (GPs), this probabilistic model maps gravitational load to a set of physiological variables, including heart rate variability (HRV), blood pressure variability (BPV), electrodermal activity (EDA), and measures of postural and locomotor behaviour. Both models were trained using a comprehensive synthesis of parabolic flight literature, enabling continuous prediction across the full spectrum of gravitational conditions from Earth gravity (1g) to microgravity (0g). By modelling neural and peripheral systems in parallel, this framework provides a holistic and quantitative approach to predicting human adaptation to gravitational perturbations, offering a tool for astronaut monitoring and countermeasure design.

Part 2 explores an experimental simulation of human awareness across multiple gravitational contexts, including microgravity (0g), Lunar gravity (0.16g), Martian gravity (0.38g), and extreme hypergravity during training or launch (2-5 g). Instead of relying on costly or ethically constrained experiments, we use an LLM to simulate a human subject's embodied awareness under these conditions. The LLM is provided with structured physical, biological, and physiological parameters describing gravitational load, autonomic responses, postural control, and sensorimotor feedback, allowing it to generate detailed descriptions of self-awareness, alertness, and conscious experience. By analysing the plausibility, consistency, and depth of these simulated introspective reports, we can explore the potential of LLMs as tools for virtual embodiment and computational phenomenology.

## Methods

### Part 1: Deep-Learning Models of Gravity-Awareness

To investigate gravity-awareness quantitatively, we developed a computational framework that integrates neural and physiological data across a range of gravitational conditions. By leveraging established findings from altered-gravity studies, particularly parabolic flights, this framework captures both central and peripheral adaptations, providing a unified approach to model how the human brain and body respond to changes in gravitational load. The framework consists of two complementary components, each designed to predict distinct aspects of the gravity-adaptive response.

**The CorticalG Model**

To investigate the continuous relationship between gravity and key neurophysiological markers, we developed a Fourier-feature-augmented multi-layer MLP for multi-output regression. This model, which we refer to as CorticalG, captures the non-linear mapping from a single gravitational input to a multi-output vector representing EEG band-power states. Rather than classifying discrete states, the model predicts the percentage change in power for specific EEG frequency bands corresponding to a given g-load, enabling a fine-grained understanding of how brain dynamics scale across gravitational conditions from microgravity to hypergravity.

**EEG Dataset Description**

We constructed a large-scale synthetic dataset to systematically model neural activity across a continuous spectrum of gravities (g-loads). The dataset consists of 94,000 simulated observations representing 20 synthetic participants, each exposed to the full range of g-load conditions. By generating continuous, participant-specific data, this approach captures the variability and dynamics of neural responses across different gravitational environments, while overcoming the logistical and ethical limitations of human spaceflight experiments. The resulting dataset provides a robust foundation for training and validating computational models of gravity-dependent neurophysiological activity.

G-load values correspond to discrete parabolic-flight anchors (0, 0.16, 0.38, 1.0, 1.6, 1.8 g), with 0.16 g and 0.38 g drawn from published partial-gravity analogue studies. These anchors serve as phase-level labels rather than continuous accelerometer measurements. During training, g enters the model as a continuous scalar, and the Fourier-feature encoding enables smooth interpolation at intermediate values. The dataset implements a within-subject design in which 20 synthetic participants each contribute responses at every anchor, minimizing inter-individual variability while preserving within-participant dynamics critical for modelling gravity adaptation.

The data generation process was informed by the empirical trends summarised in Table 1. Each synthetic EEG signal was generated by combining alpha, beta, mu, and gamma oscillatory components with realistic noise levels matched to human EEG recordings. Gravitational load was treated as a continuous predictor, and each data point was annotated with EEG band power, simulated cortical sources, and gravitational-level metadata.

This synthetic dataset serves as a training and benchmarking resource for ML models predicting neural adaptation across gravitational environments. Validation against empirical trends from microgravity and centrifuge studies confirmed that the simulated power distributions and inter-band correlations are aligned with known physiological phenomena, particularly alpha suppression and beta/gamma enhancement as gravity increased. The dataset therefore provides a controlled, reproducible foundation for testing hypotheses about human neural adaptation to gravity, while paving the way for future empirical verification using in-flight or parabolic flight EEG. The synthetic dataset of 20 participants in altered gravity conditions is available in open access: https://doi.org/10.5281/zenodo.17450642

***Table 1.*** **Foundational Literature Informing the Synthetic Dataset: Capturing EEG Rhythms and Their Modulation Across Gravitational Conditions.** *Note:* Theta and Delta waves are not included in the table, as they primarily reflect sleep and drowsiness rather than active sensorimotor or cognitive processing.

| EEG Band | Frequency Range (Hz) | Associated Brain Network / Region | Gravitational Condition | Observed Modulation / Effect | Functional Interpretation | Reference |
|---|---|---|---|---|---|---|
| **Gamma (γ)** | 30–45 Hz | Global (widespread cortical activity) | Across conditions | Low-gamma shows minimal systematic modulation | Represents global synchronization with limited gravity dependence; far less pronounced. | Cebolla et al., 2016<br>Cheron et al., 2016. |
| **Beta (β)** | 13–30 Hz | Prefrontal Cortex (PFC) | Hypergravity (1.8 g) | Increased beta activity in prefrontal regions. | Linked to elevated alertness and cognitive workload, linked in stress. | Smith et al., 2013 |
| | | | Microgravity (0g) | Increased activity. | | |
| **Mu (μ)** | 9-11 Hz | Sensorimotor Network (SMN) | Microgravity (0 g) | Increased event-related desynchronization (ERD) over sensorimotor regions | Indicates reallocation of neural resources toward postural and proprioceptive control | Quivira-Lopesino et al., 2025 |

| Alpha (α) | 8–13 Hz | Default Mode Network (DMN) | Microgravity (0 g) | Significant reduction in alpha power and functional connectivity | Reflects reduced cortical inhibition and altered baseline awareness | Pusil et al., 2023 |
|---|---|---|---|---|---|---|

**Simulation and Preprocessing of CorticalG**

The created dataset simulates a series of parabolic flights, with each flight consisting of five distinct gravitational phases. The simulation produced 64-channel EEG signals at a frequency of 250 Hz. Each 25-second gravitational phase was designed with phase-modulated, band-specific oscillatory sources. These sources were then projected onto realistic scalp topographies for the Default Mode Network (DMN), Prefrontal Cortex (PFC), and Sensorimotor Network (SMN).

In this simulation, each "run" represents a single parabolic maneuver sequence consisting of five consecutive gravitational phases (1g baseline → 1.8g pull-up → 0g microgravity → 1.8g pull-out → 1g recovery). Each of the 20 synthetic subjects completed 20 such runs. EEG band-power targets were computed as percentage change relative to the mean power of the initial 1g phase within each run. Per-parabola normalization, which could be used to capture these adaptation dynamics, should be explored in future work with empirical flight data.

We note that in actual parabolic flight operations, the pull-up phase preceding microgravity typically reaches approximately 1.8g, whereas the pull-out (recovery) phase following microgravity is more variable and averages approximately 1.5g due to less constrained aircraft maneuvering. In the present simulation, both hypergravity phases were assigned a nominal g-load of 1.8g. This simplification was adopted because the primary literature sources used for our anchor points (e.g., Cebolla et al., 2016; Cheron et al., 2016) generally report pooled hypergravity data without distinguishing pull-up from pull-out. Differentiating these phases with distinct g-labels represents a refinement for future modeling work.

The raw signals underwent a streaming-compatible processing workflow. This involved applying a band-pass filter (1-45 Hz), a notch filter (50 Hz), and normalizing each channel using a z-score. The continuous data was then segmented into overlapping 2 second windows with a 0.5 second stride. This segmentation was carefully managed to ensure that no single window crossed the boundary between different gravity phases.

Finally, the ground-truth target variables were calculated. For every window, the average power was computed for the alpha, beta, mu, and gamma frequency bands across their relevant cortical regions (DMN, PFC, SMN, and global, respectively). These power values were then normalized to represent a percentage change relative to the mean power of the initial 1g phase for that specific run. The resulting dataset contains entries that link a specific g-load to the corresponding percentage changes in the four neurophysiological markers.

The model architecture is designed for efficiency, mapping a single continuous input (g load) to four continuous outputs (band-power changes).

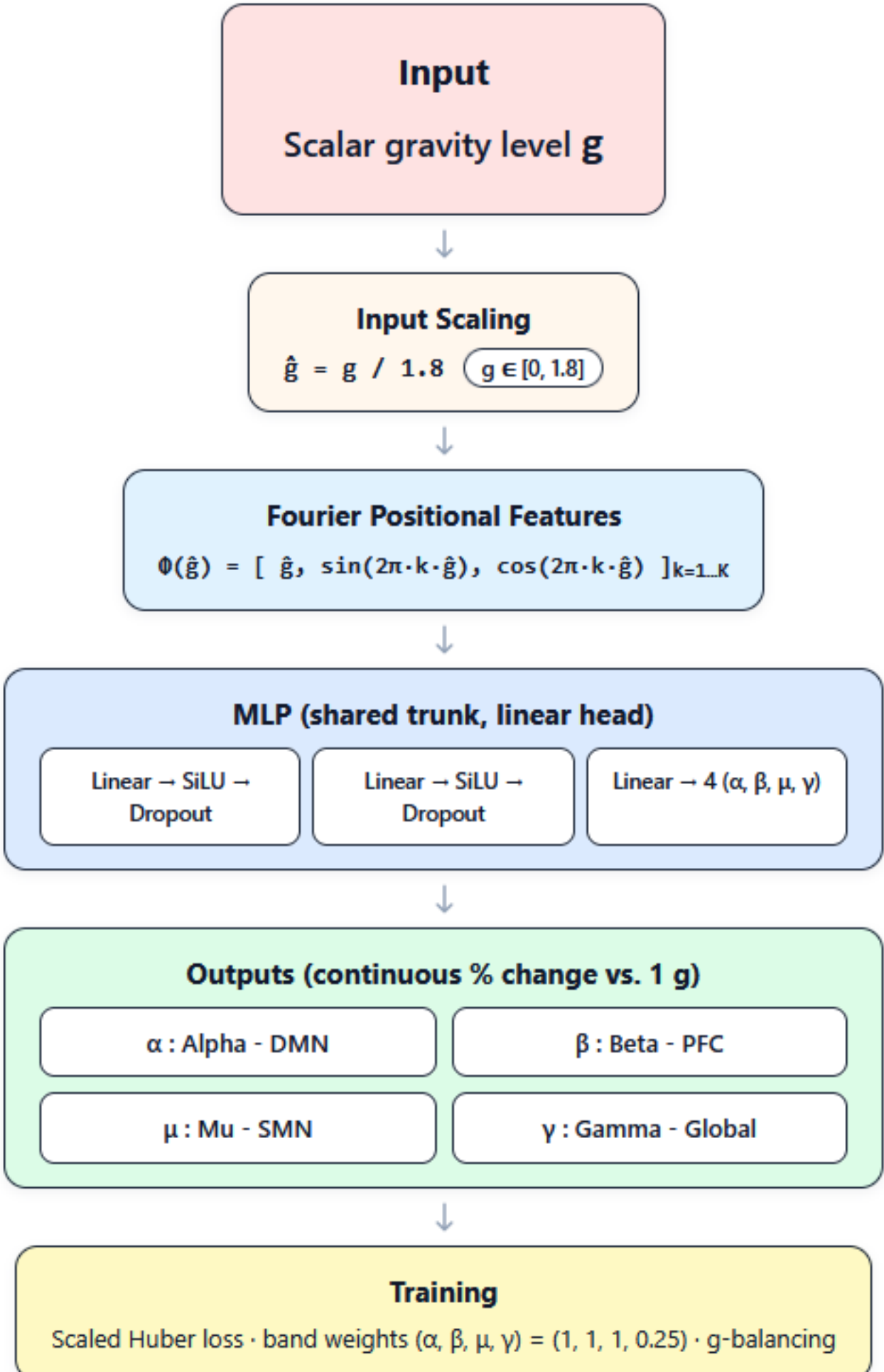

**Figure 2. Model 1 — CorticalG for Multi-Output Regression.** Schematic of the workflow used to predict relative EEG band-power changes from continuous gravity input.Input & Scaling: A scalar gravity value (g) is normalized within [0, 1.8] and anchored to empirical reference points (0, 0.16, 0.38, 1.0, 1.8 g). Fourier Features: The normalized g is expanded into sinusoidal positional encodings, enabling the model to learn smooth, periodic mappings across the g-range. MLP Architecture: A shared Multi-Layer Perceptron (two hidden layers, SiLU activation, Dropout) extracts common representations of gravity-driven EEG modulation. Outputs: A multi-output linear head predicts relative (%) changes for four bands — α (DMN), β (PFC), μ (SMN), and γ (global). Training: The model uses scaled Huber loss (band-specific weighting, optional g-balancing), optimized via AdamW with weight decay. The best model is selected by validation loss.

### Architecture Overview of CorticalG

The model takes a single gravity value as input, scaled to a standard range by dividing by 1.8. This value is then mathematically expanded into a richer nine-dimensional representation using sinusoidal functions, allowing the network to capture the complex, non-linear relationship between gravitational load and EEG activity. The core is an MLP with two hidden layers of 64 units, SiLU activation, and dropout of 0.1, followed by a linear head producing the four band-power outputs (Figure2). SiLU was selected over ReLU because it maintains non-zero gradients for negative inputs, which is advantageous for smooth continuous regression; ReLU and GELU were evaluated in preliminary experiments and yielded slightly higher validation loss (Ramachandran et al., 2017; Elfwing et al., 2018). The model was trained using a Huber loss function, weighted separately for each EEG band, with equal weight assigned to alpha, beta, and mu bands, and a lower weight (0.25) assigned to gamma, reflecting its greater variability. To prevent the model from over-fitting to the most common gravitational condition (1g), training samples were rebalanced across g-levels. The Adam optimiser with weight decay (AdamW) was used for parameter updates, and the best-performing model checkpoint was selected based on validation loss.

### The PhysioG Model

A second model, **PhysioG**, was developed to predict physiological responses across gravitational loads (Table 2). We developed a non-parametric probabilistic model based on Gaussian Processes (GPs) to model the continuous relationship between g-load and a range of physiological variables. This approach is particularly well-suited due to its ability to learn flexible, non-linear functions from sparse data points and to provide principled uncertainty estimates for its predictions. The model predicts 11 key physiological targets spanning cardiovascular, autonomic, and motor systems. Similar to the EEG model, a ground-truth dataset was required for training and validation. This dataset was constructed by extracting and digitizing data from seminal parabolic flight studies that measured physiological responses at discrete gravitational levels. The dataset was built upon "anchor points" described above, ensuring that the model's learned relationships are grounded in real-world experimental findings.

**Physiological Dataset Description**

Using a custom Python script, anchor points were interpolated using piecewise-linear functions to generate a continuous dataset spanning 0g to 1.8 g. Each of 11 physiological targets (Table 2) was modelled independently using a Gaussian Process framework, with percentage-change values computed relative to the 1g baseline to facilitate cross-feature comparison. Detailed information on synthetic data creation is available in Supplementary Material and in the open access: https://doi.org/10.5281/zenodo.17450642

***Table 2.*** Human physiological conditions as used to train the **PhysioG** model.

| Parameter | Units | Functional System / Composite | Gravitational Condition | Observed Modulation / Effect | Functional Interpretation |
| --- | --- | --- | --- | --- | --- |
| SDNN (HRV) | ms | Global heart-rate variability; aggregates slow and fast components | Across conditions (1 g, 0 g, ≈1.8 g) | **Hyper-g:** ↓ SDNN. **0 g:** ↔/↑ vs 1 g | Autonomic balance shifts sympathetic under orthostatic load; window-length sensitive |
| RMSSD (HRV) | ms | Short-term vagal HRV; respiratory sinus arrhythmia | Across conditions | **Hyper-g:** ↓ RMSSD. **0 g:** ↔/↑ vs 1 g | Vagal modulation suppressed by load; preserved/elevated in reduced-g |
| Mean NN | ms | Average R–R (inverse of HR) | Across conditions | **Hyper-g:** ↓ (tachycardia)**0 g:** (bradycardic shift) | Chronotropic change via blood redistribution and baroreflex |
| Heart Rate | bpm | Beats per minute from NN | Across conditions | **Hyper-g:** ↑ HR. **0 g:** ↔/↓ from baseline; peaks at transitions | Cardiac stimulation under orthostatic stress; unloading in 0 g |
| Systolic BP Mean | mmHg | Mean systolic arterial pressure | Across conditions | **Hyper-g:** ↑. **0 g:**↔/↓ vs 1 g | Increased peripheral resistance and ejection load in hyper-g; centralization in 0 g |
| (SBP) High-Freq Power | $mmHg^2$ | BP variability in HF (~0.15–0.40 Hz); fast baroreflex, respiratory coupling | Across conditions | Modulated by breathing rate/depth; ↑ with maneuver-related hyperventilation; more stable in steady 0 g | Marker of ventilatory–baroreflex coupling and short-latency vascular control |
| Skin Conductance — Right (SCL) | μS | Sympathetic sudomotor (tonic arousal) | Transitions and hyper-g; 0 g (novelty/stress) | **Hyper-g** (pull-up/pull-out): ↑ SCL. 0 g: possible ↑ from novelty | Arousal/attention index; sensitive to maneuver stressors |
| Skin Conductance — Left (SCL) | μS | Same construct; bilateral control | Transitions and hyper-g; 0 g | Profile mirrors right; useful for drift/artefact checks | Bilateral stability and artefact control |
| SC Asymmetry (Right–Left) | μS | Lateralized sympathetic response | Transitions | Transient deviations; magnitude/sign increase with arousal | Highlights side-dominant responses; separates global rise from local artefacts |
| Trunk Activity | count s/s | Torso accelerometry; bracing/postural effort | Transitions hyper-g; steady 0 g | Hyper-g/transitions: ↑. Stable; 0 g: ↓ | Proxy for postural muscle engagement; movement confound marker for EDA/HRV |
| Wrist Activity | **coun** s/s | Wrist accelerometer; gross upper-limb motion | Transitions; steady phases | Peaks during transitions/tasks; lower at steady baseline | Movement flag to interpret EDA/HRV; helps filter artefacts |

*Note.* Heart Rate Variability (HRV); Normal-to-Normal (sinus) beat (NN); Standard Deviation of NN (SDNN), Root Mean Square of Successive Difference (RMSSD), Systolic Blood Pressure (SBP), Skin Conductance Level/Response (SC(L)/(R)), Right minus Left (Right-Left), Electrodermal Activity (EDA), Galvanic Skin Response (GSR).

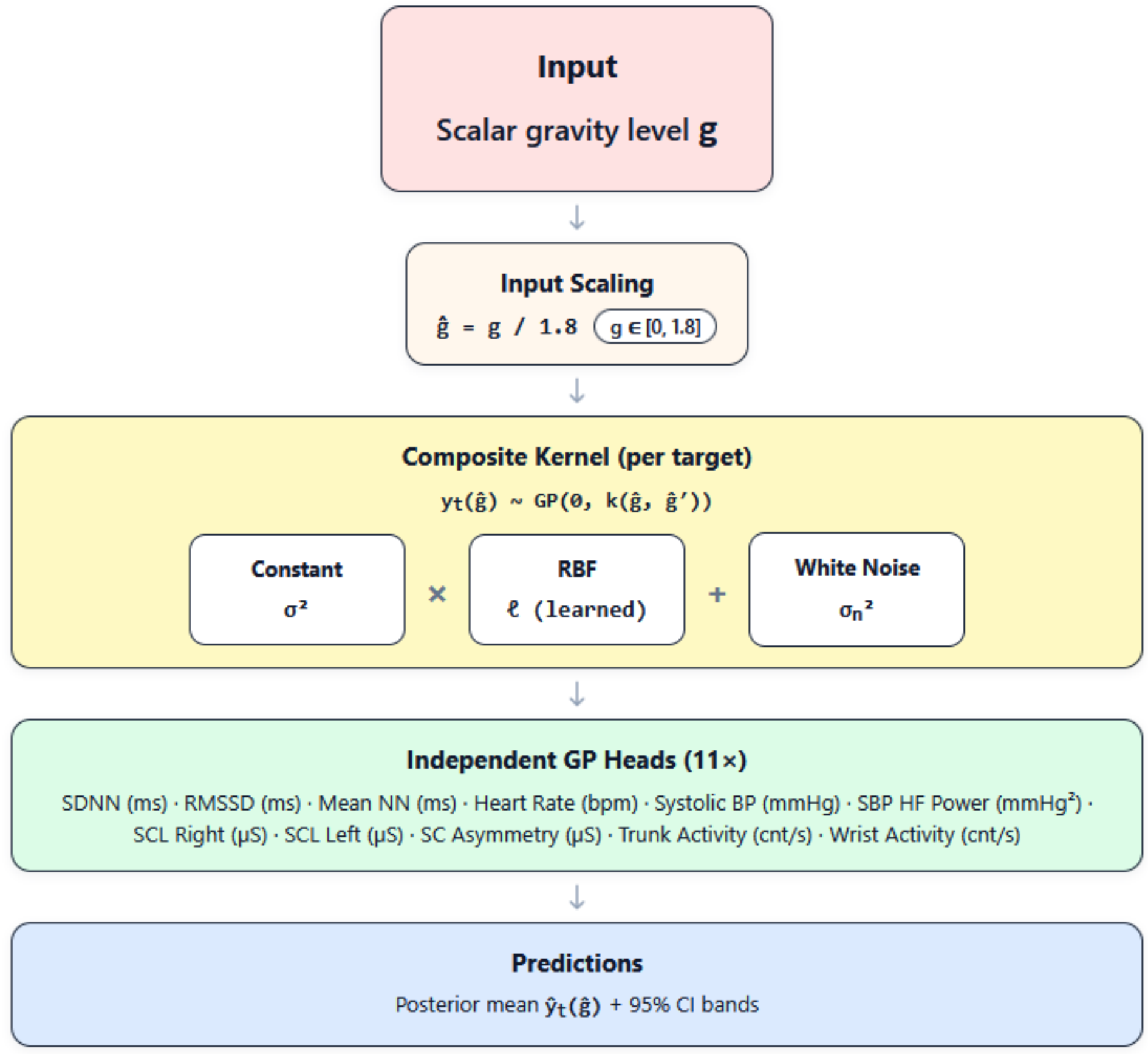

**Figure 3. Model Architecture for PhysioG – physiological Independent Gaussian Processes per target (1-D input g).** The model takes a continuous gravity input g, which is first scaled. A composite kernel, consisting of a Constant, RBF, and White Noise component, is defined for each of the 11 physiological targets. The kernel parameters are learned independently for each target by maximizing the log-marginal likelihood. This results in 11 parallel Gaussian Process "heads," each specialized for one physiological feature (e.g., SDNN, SCL, trunk activity). The models are trained and evaluated using Leave-One-Subject-Out (LOSO) cross-validation. The final output provides a posterior mean prediction and 95% confidence intervals for each target, enabling robust, uncertainty-aware reporting. The training objective per target was to maximise log-marginal likelihood on synthetic participants (LOSO); Cross-validation: Leave-One-Subject-Out (LOSO) fit per target on trained subjects: predict held-out subjects. Curves reported as mean ±1.96 SD.

### Simulation and Preprocessing of PhysioG

To model the physiological effects of varying gravitational loads, we implemented PhysioG (Figure 3). The model comprises 11 independent Gaussian Process regressors, each predicting one physiological variable (e.g., HRV, skin conductance, trunk activity; Table 2) from a single continuous gravity input. This independence allows each regressor to learn the smoothness and noise characteristics specific to its target signal. Each Gaussian Process uses a composite kernel (eq. 1):

$$k(g, g') = \sigma^2 * RBF(\ell) + \sigma n^2. \quad (1)$$

The composite kernel combines a constant term (controlling variance), an RBF term (capturing smooth non-linear g-dependence via length scale $\ell$), and a white-noise term (accounting for observation noise). The three hyperparameters ($\sigma^2$, $\ell$, $\sigma_n^2$) are learned independently per target by maximizing the log-marginal likelihood. To ensure generalization to unseen subjects, we employed Leave-One-Subject-Out (LOSO) cross-validation, training iteratively on all subjects except one held out for testing.

For constructing the physiological anchor dataset, piecewise linear interpolation was preferred over cubic splines for two reasons. First, with only 3–6 anchor points per physiological variable, unconstrained cubic splines are susceptible to oscillatory artifacts between anchors—a well-known issue with polynomial interpolation of sparse data (Runge, 1901). Such oscillations could produce physiologically implausible intermediate values (e.g., negative heart rates). Second, the downstream Gaussian Process models learn their own smooth interpolation via the RBF kernel, making the anchor dataset's role primarily one of providing ground-truth reference values rather than smooth curves. We note that the detailed supplementary formulation employs the monotone cubic Hermite interpolant (Fritsch & Carlson, 1980), which preserves monotonicity and avoids oscillation for targets where this property is appropriate.

## Results

### CorticalG and PhysioG Outputs

The trained CorticalG and PhysioG model demonstrated high accuracy and neurophysiological plausibility in predicting g-load-dependent EEG states. The model's performance was evaluated on a held-out test set, with Mean Absolute Error (MAE) serving as the primary metric to quantify the average predictive error in percentage points. The model achieved low MAE values across all four predicted EEG bands, indicating a strong predictive capability (Table 3). The final test loss, calculated using the scaled and weighted Huber objective, was 0.2315, confirming robust generalization to unseen data.

*Table 3. Model Performance on Test Set*

| Predicted EEG State | Mean Absolute Error (MAE) [%] |
|---|---|
| **Alpha – DMN** | **2.87** |
| **Beta –PFC** | **3.15** |
| **Mu – SMN** | **2.54** |
| **Gamma** | **1.98** |

*Note: MAE represents the average absolute difference between the CorticalG model predicted percentage change and the ground-truth percentage change.*

To visualize the learned relationship between g-load and brain EEG dynamics, we performed the model inference across a continuous range of gravity levels from 0g to 1.8g. Figure 4 illustrates CorticalG predictions for each of the four target bands, revealing complex, non-linear relationships that align with established neurophysiological findings.

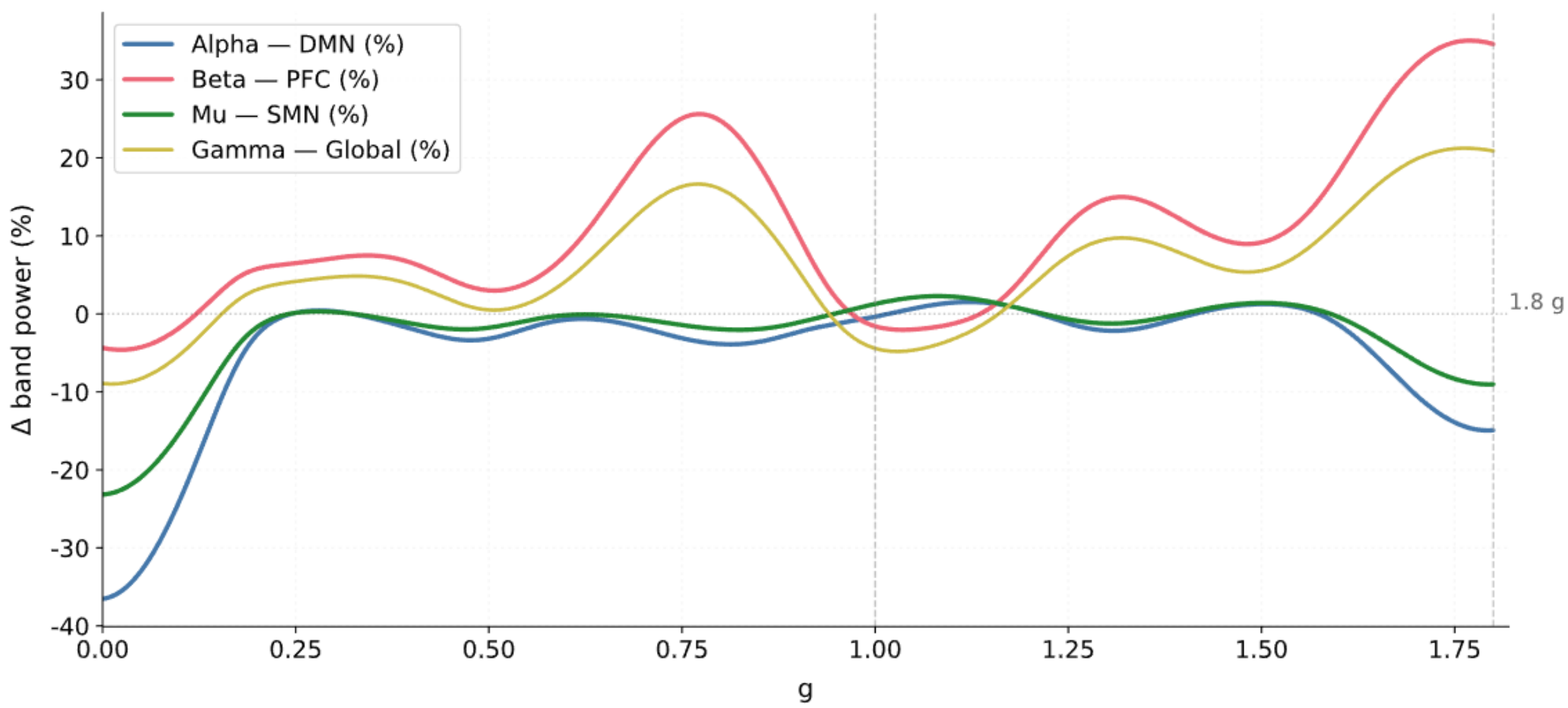

Figure 4. CorticalG predicted EEG band-power changes across continuous gravitational loads. Percentage change relative to the 1.0 g baseline for Alpha–DMN (blue), Beta–PFC (orange), Mu–SMN (green), and Gamma–Global (yellow) across g ∈ [0, 1.8]. The dashed vertical line marks Earth gravity (1.0 g); training anchors are 0, 0.16, 0.38, 1.0, 1.6, and 1.8 g. Anchor-supported effects. Alpha and mu show maximal suppression in microgravity (≈ −37 % and −23 % at 0.0 g), reproducing the DMN alpha reduction and sensorimotor mu reallocation reported during weightlessness (Pusil et al., 2023; Cebolla et al., 2016). Beta and gamma rise monotonically toward hypergravity, peaking at ≈ +35 % and +22 % at 1.8 g, consistent with cortical arousal and cognitive-load signatures of high-g exposure (Smith et al., 2013). All curves cross zero at 1.0 g, confirming correct anchor calibration. Interpolation between sparse anchors (0.75–1.75 g region). The local secondary peak in Beta and Gamma near 0.7 g and the corresponding inflection between 1.25–1.75 g are not directly observed physiological effects. They arise from the Fourier-feature MLP interpolating across the two largest anchor gaps in the training data — between 0.38 g (Mars) and 1.0 g (Earth), and between 1.0 g and 1.6 g — where no empirical reference exists. These intermediate-range predictions are model extrapolations and require validation in partial-gravity protocols.

CorticalG predictions revealed distinct, band-specific responses to g-load (Figure 4). Alpha (DMN) and mu (SMN) showed strong inverse scaling with gravity, with maximal suppression at 0g; beta (PFC) and gamma scaled positively with g-load, peaking near 1.8 g. All curves crossed zero at the 1g calibration anchor. Intermediate-range model behaviour is discussed in the Figure 4 caption above.

**Peripheral and Autonomic Dynamics (PhysioG)**

The PhysioG model provided robust, uncertainty-aware predictions for 11 key physiological variables (Figure 5; Table 2 above). The model learned several key systemic responses. HRV and BPV (Figure 5, Top Panel) predictions show a clear signature of autonomic adaptation. In microgravity (0.0g), there is a marked increase in vagal tone, evidenced by the significant positive peaks in SDNN and RMSSD and a corresponding drop in heart rate. Hypergravity suppresses these metrics, indicating sympathetic dominance. Notably, HF SBP, a proxy for baroreflex sensitivity, shows a dramatic decrease of over 60% in microgravity.

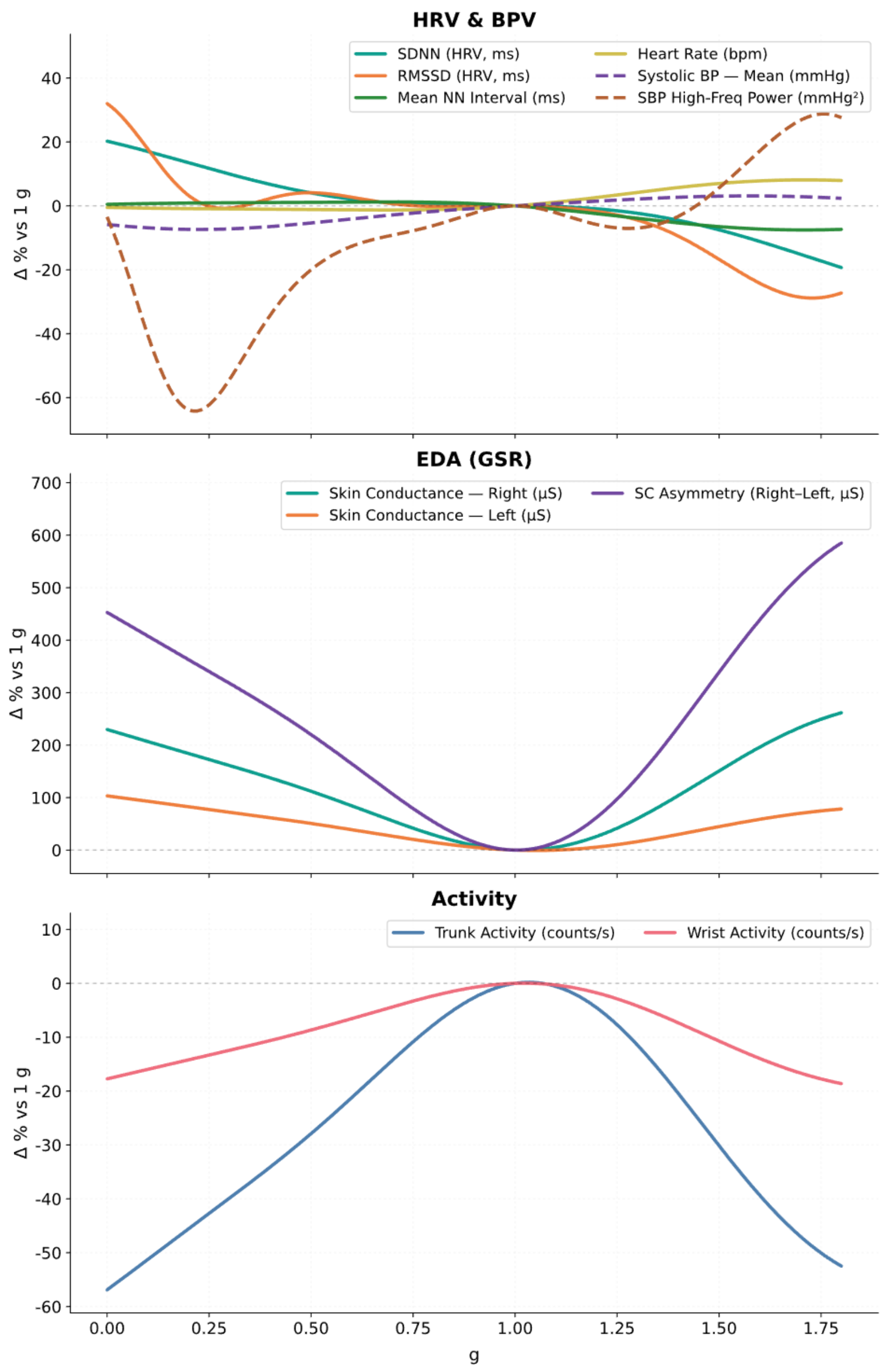

**Figure 5. Predicted Physiological States Across Continuous G-Loads in 11 variables described by PhysioG.** The figure displays the model's posterior mean predictions for (Top) Heart Rate Variability and Blood Pressure Variability, (Middle) Electrodermal Activity, and (Bottom) Physical Activity. Predictions are shown as a percentage change relative to the 1.0g baseline across a continuous gravity spectrum.

Electrodermal activity (Figure 5, Middle) showed a V-shaped response: skin conductance was lowest at 1.0g and increased sharply at both extremes (right SCL > +200% at 1.8g), indicating high sympathetic arousal at both gravitational extremes. Physical activity (Figure 7, Bottom) showed an inverted U-shape: trunk activity peaked at 1.0g and dropped > 50% at 0.0g, reflecting postural muscle unloading, while wrist activity remained relatively stable. Gravity extremes thus drive opposing autonomic (↑) and postural (↓) adjustments.

**Part 2: LLM Simulation of Gravity-Awareness**

To complement the quantitative models, we used a LLM (Claude 3.5 Sonnet) to generate phenomenological first-person descriptions of awareness across gravitational conditions. This approach is motivated by the recognised potential of LLMs to function as *in silico* participants: generating structured, cognitively plausible responses to scenarios that are ethically constrained, logistically costly, or physically inaccessible to human experimentation (Aher et al., 2023; Argyle et al., 2023). Rather than replacing empirical data, the simulation serves as a theoretically grounded *thought experiment*: by holding physiological parameters constant and varying only gravitational load, we isolate the predicted phenomenological signature of each gravitational condition independent of individual variability. The model was prompted with a fixed set of physical and physiological parameters drawn from the altered-gravity literature (EMG 20–60%, HRV 20–100 ms, postural sway 5–10 mm, stride length 1.2–1.5 m, stride frequency 1.1–1.4 Hz, linear acceleration scaled to g-level, head tilt 0°) and asked to report on alertness and self-awareness under each condition. No other individual-participant characteristics were specified, ensuring that differences in output across conditions reflect gravitational load alone.

## Methods

### Prompt Engineering Protocol

Claude 3.5 Sonnet was selected as the LLM for this simulation following informal pre-testing in which we compared outputs from Claude 3.5 Sonnet, ChatGPT-4o, DeepSeek-V3, and Gemini 1.5 Pro using identical prompts. Claude's responses were subjectively judged by the authors to describe self-awareness and embodied experience most cohesively and with the greatest phenomenological detail. This choice is further supported by emerging evidence that Claude models produce narrative outputs judged to be particularly coherent, well-structured, and human-like in qualitative research contexts (Jenner et al., 2025) .

To probe the embodied experience of awareness across gravitational environments, the model was instructed to engage in a structured mental simulation exercise. In this scenario, a participant was asked to imagine themselves as a human defined by a series of physiological and biomechanical parameters influenced by gravitational load. These included: linear acceleration (la), centered around 9.81 $m/s^2$ and scaled according to gravitational condition; head tilt (ht), ranging from –15° to 15°; heart rate variability (hrv), from 20 to 100 milliseconds; muscle activation measured via electromyography (emg), scaled from 20% to 60% according to gravitational demand; postural sway (ps), ranging from 5 to 10 millimeters; stride length (sl), from 1.2 to 1.5 meters; and stride frequency (sf), from 1.1 to 1.4 Hz, both parameters adjusted to reflect biomechanical adaptations across gravitational levels. Time was normalized using a standard temporal index to control for perception of duration across conditions. These parameters reflect standard variables recorded in altered-gravity research, including HRV (Migeotte et al., 2003; Clément et al., 2020), manual control under altered g (Rosenberg et al., 2018) , head tilt (Shelhamer, 2016), gait metrics (Clément et al., 2024), and postural sway (Hardy et al., 2025). The key difference from human participants is that they experience these states bodily rather than as numerical input — a limitation discussed below.

The LLM was presented with six gravitational scenarios, microgravity (0g), lunar (0.16g), Martian (0.38g), Earth (1g), spacecraft take-off (4g), and peak re-entry (6g), and asked to self-report alertness and self-awareness in each. Alertness and self-awareness were defined following Posner & Petersen (1990), Morin (2011), and Oyserman et al. (2012) as described in the Introduction (Gravity-Awareness framework section), with alertness stratified into six levels (unconscious, very drowsy, slightly drowsy, normal, highly alert, panic) and self-awareness as absent/present. The full prompt is provided in Supplementary Materials (Section S2).

## Results

### LLM Simulation Output

The response from Claude 3.5 Sonnet provided a series of detailed, plausible first-person narratives capturing variations in cognitive state and physiological awareness across gravitational states. The generated texts highlighted distinct profiles, for example, feelings of spatial disorientation and sensory fragmentation in 0g, heightened motor effort and strategic bodily adjustments in Martian gravity, and intense physiological strain with narrowing of consciousness during high-g events. Representative excerpts from Claude 3.5 Sonnet's responses are shown below; full responses for all six conditions are provided in Supplementary Materials (Section S3).

Across the six conditions, Claude's responses show a coherent gradient: microgravity elicits heightened alertness with disorientation, partial gravity increases attentional demand for motor coordination, and extreme hypergravity narrows self-awareness under acute physiological strain. Earth gravity serves as the optimally calibrated baseline.

The LLM's characterization of postural parameters as "irrelevant" during high-g conditions requires critical qualification. While occupants are indeed physically restrained during spacecraft launch and re-entry, postural control becomes biomechanically more - not less - significant under elevated g-loads. A given postural deviation $\Delta x$ produces a gravitational torque proportional to $g \times \Delta x \times$ body mass; consequently, the same displacement at 4g generates four times the destabilizing torque compared to 1g. For unrestrained operators such as fighter pilots executing high-g maneuvers, this effect is directly relevant to G-tolerance and loss-of-consciousness risk (Whinnery & Forster, 2013). The LLM's output reflects a simplified scenario (restrained astronaut) that does not generalize to all high-g contexts, and the omission of torque-scaling effects represents a limitation of phenomenological simulation that lacks physical modeling.

## Discussion

This study introduces a dual computational framework designed to quantitatively model the human brain's response across a continuous spectrum of gravitational loads, providing a novel lens to investigate how gravity shapes human behaviour. By integrating a multilayer perceptron (MLP) model for central nervous system dynamics with an EEG Fourier-based representation, alongside a PhysioG model for peripheral physiological states, we present a cohesive, multi-system perspective of human awareness in altered gravity. This integrative approach synthesizes neural, autonomic, and motor components into a unified framework, bridging gaps between central and peripheral responses. Our findings align closely with the empirical literature on EEG and physiological outcomes in altered gravity, revealing complementary patterns of neurophysiological adaptation and autonomic regulation, where cortical dynamics mirror predictable shifts in cardiovascular, electrodermal, and motor activity.

**Predictive Models of Gravity-Awareness**

The CorticalG model effectively learned the non-linear relationships between gravitational load and key neural oscillations. Suppression of alpha and mu rhythms in microgravity aligns with established findings, reflecting both altered resting-state cognition - particularly within the default mode network (DMN) - and a fundamental reorganization of sensorimotor strategy due to postural unloading (Cebolla et al., 2016; Pusil et al., 2023). Conversely, enhanced beta and gamma power in hypergravity captures the cortical signature of heightened arousal and cognitive stress in hypergravity environments (Smith et al., 2013). These central predictions are reinforced by the PhysioG model, which uncovers a V-shaped pattern of electrodermal activity (EDA/GSR): high sympathetic arousal emerges at both microgravity (0g) and hypergravity (1.8g), whereas the baseline terrestrial gravity environment reflects homeostatic stability. This indicates that the autonomic nervous system interprets both extremes as salient stressors: novel and disorienting in microgravity and physically demanding in hypergravity. Cardiovascular predictions further support this, showing increased vagal tone (RMSSD) in microgravity, consistent with early-stage cardiovascular deconditioning observed in parabolic flight studies (Widjaja et al., 2015). Peripheral motor patterns complement this picture: reduced trunk activity versus stable wrist motion provides a direct correlate for sensorimotor reorganization, aligning with the EEG mu-rhythm suppression indicative of altered postural strategies (Wang et al., 2025).

Two distinct trends emerge across peripheral metrics. First, electrodermal activity demonstrates a clear V-shaped response: skin conductance is minimal at 1.0 g but increases sharply in micro- and hypergravity, with right-hand conductance rising over 200% at 1.8 g, highlighting pronounced sympathetic engagement (Amin & Faghih, 2022). Second, body activity follows an inverted U-shaped pattern: trunk motion peaks near 1g and decreases by more than 50% under microgravity due to axial unloading, while wrist activity remains relatively invariant, suggesting limb motion is less dependent on gravitational load than core stabilization.

Methodologically, this study demonstrates the strength of a literature-informed, data-synthesis approach. Anchoring models in robust empirical findings allows interpolation across a continuous range of g-loads despite the scarcity of continuous experimental data. The use of two complementary modeling architectures was deliberate: the lightweight CorticalG captures complex EEG response curves efficiently, while the Gaussian Process probabilistic model provides essential uncertainty estimates for noisier peripheral signals. The dual-model outputs are consistent and mutually reinforcing. High-arousal states inferred from the EDA spikes at 0 g and 1.8 g correspond with EEG predictions of elevated beta/gamma activity in hypergravity and alpha suppression in microgravity. Simultaneously, EEG-indicated central sensorimotor adaptations (mu-rhythm suppression) are mirrored by reduced trunk activity, confirming a fundamental shift in motor control strategy under changing postural demands. This dual-model framework offers a robust, multi-system method to quantify continuous human responses to gravitational variation. However, it is important to note the limits of this approach. Both models are predictive and interpolative, learning mappings between g-load and observable physiological outputs from literature-derived data. They generalize these mappings to intermediate g-values but do not provide mechanistic explanations for why alpha rhythms are suppressed in microgravity or why sympathetic arousal forms a V-shaped curve. The distinction between predictive accuracy and mechanistic insight is fundamental (Marr, 1982; Friston, 2010; Angelaki & Cullen, 2008). These models should therefore be interpreted as quantitative monitoring tools for multi-system signatures of gravity-awareness, rather than explanatory frameworks for the underlying neurobiological mechanisms.

**LLM as Gravity-on-Awareness Simulator**

To complement the quantitative models, we implemented a structured prompt in Claude 3.5 Sonnet to simulate subjective human experiences across six gravitational environments: microgravity (0 g), partial gravity (Moon: 0.16 g; Mars: 0.38 g), and hypergravity (4-6 g, corresponding to spacecraft take-off and re-entry). The prompt specified first-person physiological and biomechanical parameters—heart rate variability, postural stability, vestibular load, EMG activation, stride metrics, and head tilt—alongside cognitive descriptors such as alertness and bodily self-awareness. It is crucial to interpret these outputs with caution. The LLM translates objective numerical parameters into phenomenological descriptions, whereas human perception of these parameters is inherently subjective. This limits direct comparability with self-report data from parabolic flights or centrifuge experiments (Lush et al., 2021). Consequently, the LLM simulation functions as a theoretically grounded, complementary thought experiment rather than a substitute for empirical data. Future work should develop validated verbal descriptors or ecological vignettes based on existing spaceflight self-report instruments (e.g., Space Motion Sickness scale) to enable meaningful LLM-human phenomenological comparisons.

Despite these caveats, Claude's outputs aligned closely with the dual-model predictions. In 0 g, the model reported disorientation and a "floating, untethered" bodily awareness, consistent with EEG-predicted DMN alpha suppression and PhysioG evidence of elevated vagal tone and trunk unloading. In hypergravity (1.8 g), the LLM described narrowed self-awareness and intense physiological strain, matching increased prefrontal beta/gamma activity and EDA spikes predicted by the models. This cross-modal concordance supports the utility of LLMs as tools for simulating embodied awareness in environments that are experimentally challenging or inaccessible. Future iterations could integrate individualized physiological baselines to enable subject-specific prediction of gravitational stress susceptibility, including risk of G-induced Loss of Consciousness (G-LOC).

The LLM's prediction of an increased stride length (~1.8m) under lunar gravity warrants careful interpretation. At 0.16g, stride lengths exceeding ~1.5m would result in loss of ground contact, transitioning locomotion from walking to bounding and slowing in movement —a qualitative shift the LLM does not capture —which is, in fact, consistent with locomotion patterns observed during Apollo surface activities (e.g., https://youtu.be/S9HdPi9Ikhk?si=jGRbS_HBlBoaNtle&t=3183) and predicted by biomechanical models of reduced-gravity locomotion (Minetti, 2001; Carr & McGee, 2009). While the LLM did not explicitly label this as a gait transition, the predicted parameter values are consistent with the shift from walking to bounding that characterizes actual lunar locomotion. And, we must acknowledge that the gait is indeed could be also considered slower. Nonetheless, the model's failure to flag this as a qualitative change in locomotion mode—rather than a simple parametric increase—illustrates a limitation of using continuous numerical parameters to describe phenomena that involve discrete biomechanical regime changes.

**Astronaut Training and Practical Implications**

One immediate translational application lies in enhancing astronaut training. The CorticalG's efficiency enables real-time monitoring of awareness-related neural patterns during high-G centrifuge or VR-based gravity simulations, while the GP physiological model tracks multi-system responses with uncertainty estimation. This dual-mode system facilitates acute intervention and longitudinal monitoring. For example, VR simulations could dynamically adapt task difficulty in response to predicted cognitive strain or autonomic arousal, tailoring training to individual neurophysiological profiles (Varandas et al., 2022).

In practice, the framework enables quantification of gravity-awareness across neurophysiological, cognitive-perceptual, and sensorimotor-behavioral domains. Personalized pipelines could include: (i) real-time EEG/GP-based monitoring to build individual gravity-adaptation schedule; (ii) closed-loop VR or centrifuge exposures titrated to target weak points in sensory weighting or postural control; (iii) neurofeedback and task-level coaching when biomarkers signal elevated risk; (iv) curriculum-based meta-learning to optimize adaptation and retention; and (v) optimal-control metrics to set mastery criteria. Such an approach can accelerate adaptation, increase G-tolerance, reduce motion sickness, and preserve executive function—critical for long-duration, high-performance space missions.

Integration with LLM simulations further enables predictive insight into subjective awareness degradation, supporting pre-flight screening, adaptive training, and cognitive self-monitoring to mitigate risks of G-LOC or microgravity disorientation.

**Limitations and Future Directions**

Several limitations warrant consideration. First, the models are trained on literature-derived anchors rather than continuous real-world physiological datasets. While providing a robust proof-of-concept, empirical validation using parabolic flight, centrifuge, or spaceflight data is essential to assess predictive robustness. Second, the LLM simulations were based on limited prompts; future studies should extend these to longer-duration microgravity exposures and more diverse scenarios. Additionally, the opacity of LLM training data constrains interpretability, as the extent to which the model encodes altered-gravity physiology is unknown (Bommasani et al., 2021). Comparative studies across different LLMs (e.g., GPT-4, Gemini, DeepSeek) would help establish robustness and model-specific biases (Zheng et al., 2023).

Another limitation is the models' focus on g-load as the sole independent variable, neglecting psychological stress, task demand, and individual differences that interact with gravitational adaptation. Future work should incorporate these dimensions to build more comprehensive, personalized models. Likewise, our current dataset is anchored to $g \leq 1.8$, limited in capturing high-G thresholds that precipitate G-LOC. Including data from 2 g and higher centrifuge experiments would enable modeling of non-linear physiological cascades leading to cahnge or loss of awareness (Erdem & Demir, 2006; Singh & Bhasin, 2014). Extending models to negative effective g-loads (e.g., −1 g, full-body inversion) also represents a promising avenue, given potential physiological analogies to microgravity adaptations (Arbeille et al., 2001).

We also aknowledge that the per-run normalization strategy of the synthetic dataset does not account for within-session adaptation effects: in real parabolic flights, the first parabola typically elicits stronger physiological responses than later parabolas due to habituation and progressive sensory reweighting (Clément & Reschke, 2008).

Ultimately, the framework provides a strong foundation for integrating multi-modal physiological data (EEG, HRV, EDA, motion tracking) to generate personalized predictions of embodied cognition and awareness, supporting adaptive biofeedback, resilience training, and situational preparedness in spaceflight contexts despite the limitations.

## Conclusion

The combination of CorticalG and PhysioG models, complemented by qualitative LLM simulations, provides a comprehensive, multi-system perspective on gravitational impacts on human cognition and physiology. The dual-model framework successfully captures continuous, non-linear effects of g-load across central and peripheral systems, including hallmark signatures such as DMN alpha and mu suppression in microgravity, enhanced prefrontal beta/gamma activity in hypergravity, and the V-shaped pattern of sympathetic arousal. This work establishes a robust computational foundation for predictive, personalized tools to monitor, simulate, and train human performance under extreme gravitational conditions, advancing both the scientific understanding of gravitational adaptation and practical approaches to astronaut health and operational readiness.

## Supplementary Material

### S1: Full LLM Prompt

*"Imagine you are a human with following parameters:*

*Physics + biology= PHYSICS (Gravity)*

*Space - 0g Earth - 1g Moon - 0.16 Mars - 0.38g Take-off - 4g Re-entry - 6g*

*BIOLOGY: 1. Linear acceleration (la): Centered around 9.81 m/ (or scaled according to the gravitational condition)*

*2. Head tilt (ht): Ranging from -15° to 15°*

*3. Heart rate variability (hrv):20-100 ms*

*4. Muscle activation/EMG (emg): 20-60% (scaled with gravity)*

*5. Postural sway (ps): 5-10 mm (scaled with gravity)*

*6. Stride length (sl): 1.2-1.5 m (derived from gravity-dependent scaling)*

*7. Stride frequency (sf): 1.1-1.4 Hz (derived from gravity-dependent scaling);*

*8. Time: A normalized temporal index*

*Given that you can be in an alertness state (defined as "Alertness is a state that enables an individual to process sensory information, maintain stability, and respond effectively to environmental demands (Posner, Petersen, 1990): Panic = Unstable in an emergency , Highly alert = Stable in an emergency , Normal alert = Optimally stable , Slightly drowsy = Unstable, Very drowsy = Critically unstable, Unconscious = No control + self-awareness (defined as "Self-awareness represents the capacity of becoming the object of one's own attention "(Morin, 2011) and "Self-identity refers to an individual's self-concept that emerges from reflective self-awareness, integrating personal history, social roles, and cultural context to create a coherent sense of self across time and situations" (Oyserman, et al. 2012); levels: absent-present). Provide*

*a short description of your state and self-awareness level for every gravitational appearance (g, described above)."*

S2 LLM answers

*"I'll analyze each gravitational scenario from a first-person perspective, considering both alertness and self-awareness states while accounting for the biological parameters:*

*Space (0g): My linear acceleration feels absent, and my postural control systems are struggling to find reference points. Heart rate variability is elevated (80-90ms) due to fluid shifts. I feel highly alert but with some disorientation - my self-awareness is present but altered, as my body schema needs to adjust to the novel environment. EMG shows minimal activation as muscles work differently in microgravity.*

*Earth (1g): All biological parameters are at baseline. Linear acceleration matches standard gravity, heart rate variability is optimal (50-60ms), and postural sway is within normal range (7mm). I'm in a normal alert state with complete self-awareness present - this is my reference environment where all systems are calibrated optimally.*

*Moon (0.16g): Stride length increases to about 1.8m due to reduced gravity, while EMG shows reduced activation (around 30%). I'm highly alert due to the novel environment, with present self-awareness but enhanced attention to movement patterns. Postural sway increases slightly as I adapt to the reduced gravitational field.*

*Mars (0.38g): Similar to lunar conditions but more stable. Stride length is about 1.6m, EMG at 40%. Normal alert state with present self-awareness, as the gravity is sufficient to maintain*

*familiar movement patterns while requiring less energy. Heart rate variability remains stable around 55ms.*

*Take-off (4g): High linear acceleration causes significant muscle strain (EMG at maximum). Heart rate variability decreases to 30ms due to stress response. I'm highly alert with present self-awareness, though focused primarily on physiological management. Postural parameters are irrelevant as I'm secured in position.*

*Re-entry (6g): Extreme physical stress with EMG maximal, heart rate variability low (20ms). I'm maintaining high alertness but approaching physiological limits. Self-awareness is present but narrowed, focusing mainly on breathing and consciousness maintenance. Movement parameters are irrelevant due to the confined position and extreme forces."*

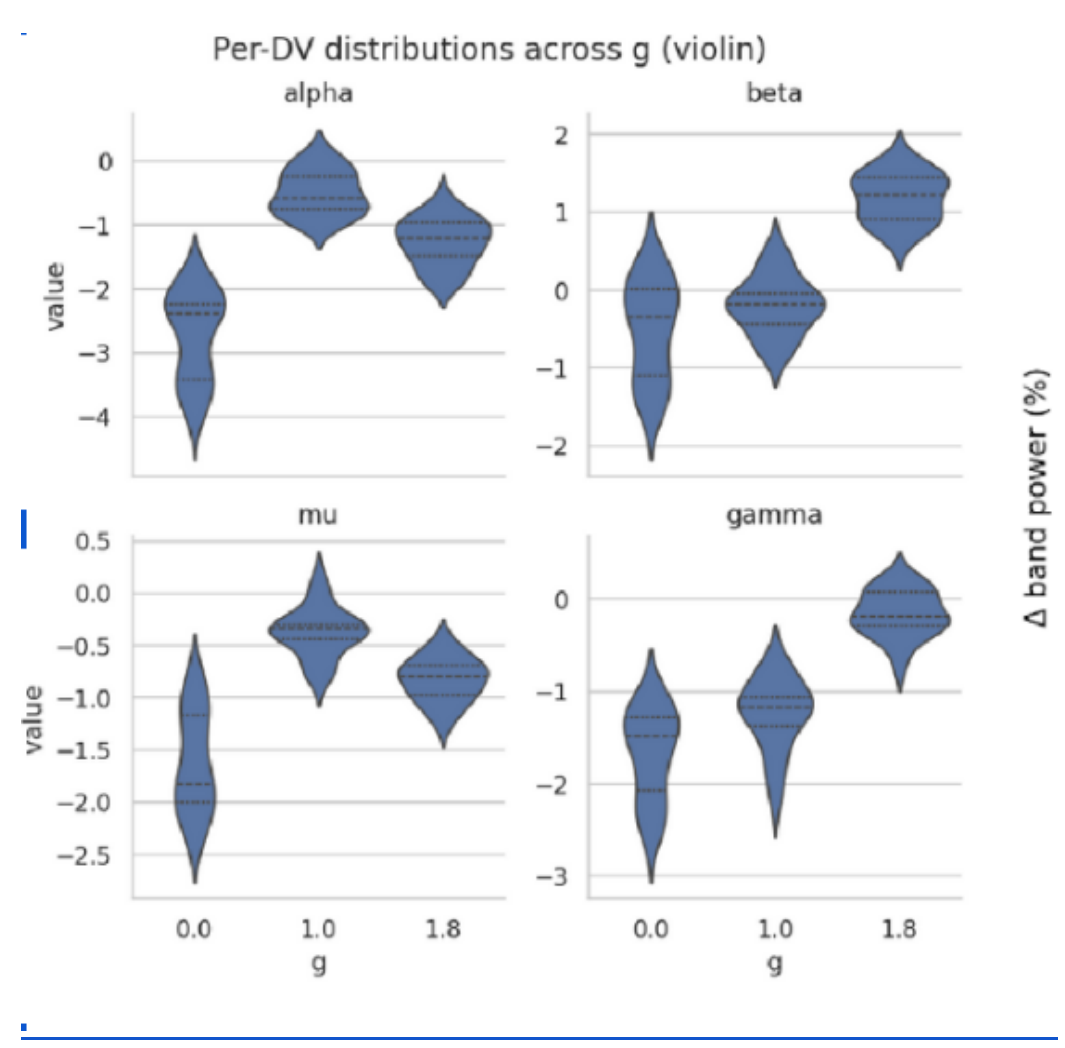